# Quantum decoherence of neutrino mass states

**Konstantin Stankevich**[a,*] **and Alexander Studenikin**[a]

[a]*Faculty of Physics, Lomonosov Moscow State University,*
*Moscow 119991, Russia*
*E-mail:* kl.stankevich@physics.msu.ru, studenik@srd.sinp.msu.ru,
a-studenik@yandex.ru

We study the interplay of the neutrino quantum decoherence and bipolar collective neutrino oscillations. Using the numerical simulation of the neutrino evolution in a supernova environment we show the suppression of the bipolar collective neutrino oscillations by the quantum decoherence of the neutrino mass states.

---

[*]Speaker

The neutrino quantum decoherence is a process of the coherent superposition violation of neutrino states that is engendered by the interaction with an external environment (see [1] and references therein). In our previous work [2] we derived the new conditions of the existence of the bipolar collective neutrino oscillations accounting for the effect of quantum decoherence of neutrino mass states.

In the present work we provide the numerical simulation for the neutrino evolution in supernova environment accounting for the interplay of two effects: (i) the neutrino quantum decoherence and (ii) bipolar collective neutrino oscillations.

The evolution of the neutrino and antineutrino fluxes accounting for the neutrino quantum decoherence and collective neutrino oscillations is determined by the Lindblad master equations

$$i\frac{d\rho_f}{dt} = [H, \rho_f] + D[\rho_f], \tag{1}$$

$$i\frac{d\bar{\rho}_f}{dt} = [\bar{H}, \bar{\rho}_f] + D[\bar{\rho}_f], \tag{2}$$

where $H = H_V + H_m + H_{\nu\nu}$ is the neutrino Hamiltonian, that accounts for the vacuum Hamiltonian $H_V$, the Hamiltonian $H_m$ of neutrino interactions with an external matter (electrons, neutrons and protons) and the Hamiltonian $H_{\nu\nu}$ of the neutrino-neutrino interaction. The density matrices $\rho_f$ and $\bar{\rho}_f$ describe evolutions the neutrino and antineutrino fluxes correspondingly. The effect of quantum decoherence of neutrino mass states is defined by the dissipator

$$D[\rho_f] = \frac{1}{2}\sum_{k=1}^{N^2-1}\left[V_k, \rho_f V_k^\dagger\right] + \left[V_k \rho_f, V_k^\dagger\right], \tag{3}$$

where $V_k$ are the Lindblad operators and $N$ is the dimension of the $\rho_f$-space ($N = 2$ for the two-flavour neutrino oscillations and $N = 3$ for the three-flavour neutrino oscillations).

In the present work we consider two-flavour neutrino approximation. Then the operators $V_k$, $\rho_f$ and $H$ can be expanded over the Pauli matrices $O = a_\mu \sigma_\mu$, where $\sigma_\mu$ are composed by an identity matrix and the Pauli matrices. In this case equation (3) can be written in the following form

$$\frac{\partial P_k(t)}{\partial t}\sigma_k = 2\epsilon_{ijk}H_i P_j(t)\sigma_k + D_{kl}P_l(t)\sigma_k, \tag{4}$$

where the matrix $D_{ll} = -diag\{\Gamma_1, \Gamma_1, 0\}$ and $\Gamma_1$ is the parameter that describes the decoherence effect of neutrino mass states.

We use a simplified two flavor monoenergetic model of supernova neutrinos [3, 4] for the numerical simulations. Within this model the expressions for the neutrino Hamiltonians in a flavour basis are given by

$$H_{vac} = \frac{\delta m^2}{4E}\begin{pmatrix} -\cos 2\theta & \sin 2\theta \\ \sin 2\theta & \cos 2\theta \end{pmatrix}, \tag{5}$$

$$H_M = \frac{\sqrt{2}}{2}G_F n_e \begin{pmatrix} 1 & 0 \\ 0 & -1 \end{pmatrix}, \tag{6}$$

$$H_{\nu\nu} = \sqrt{2}G_F n_\nu \left((1+\beta)\rho_f - \alpha(1+\bar{\beta})\bar{\rho}_f\right), \tag{7}$$

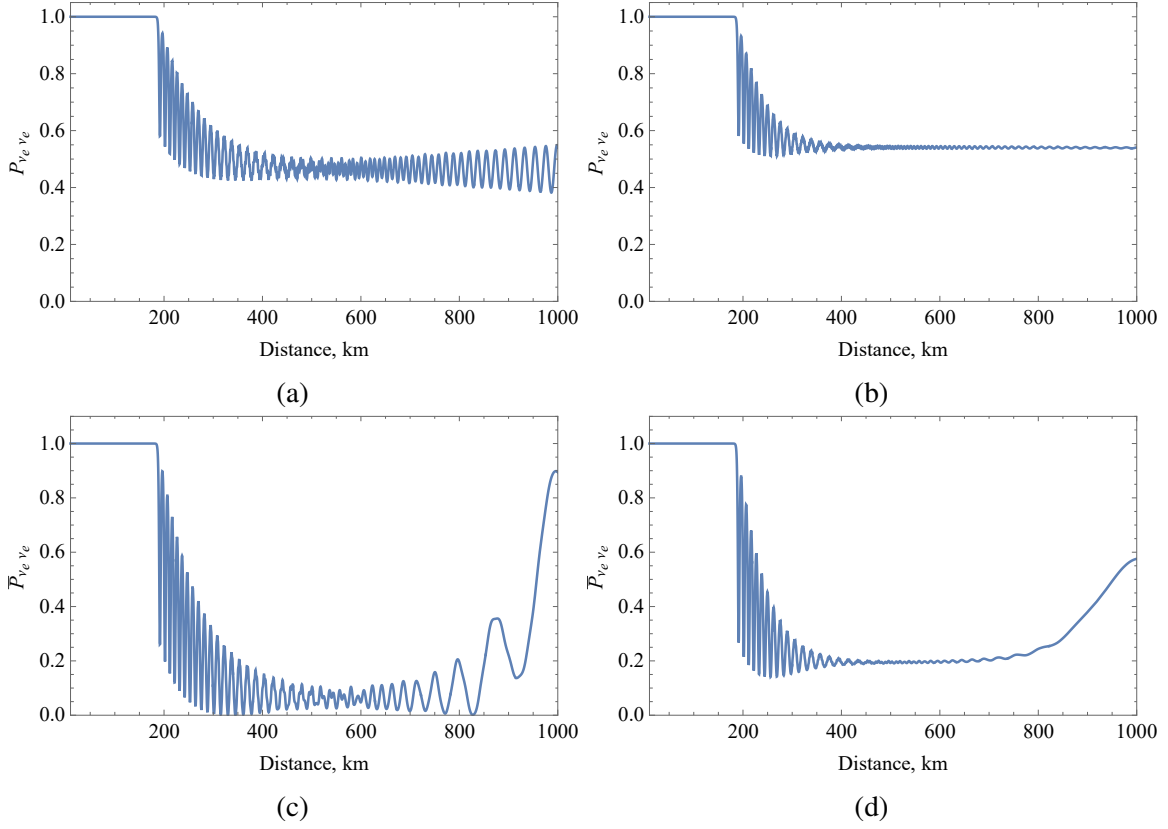

**Figure 1:** figure (a) shows the survival probability of an electron neutrino with zero effect of neutrino quantum decoherence and figure (b) shows the survival probability of an electron neutrino accounting for the neutrino quantum decoherence ($\Gamma_1 = 10^{-21}$ GeV); figure (c) shows the survival probability of an electron antineutrino with zero effect of neutrino quantum decoherence, figure (d) shows the survival probability of an electron antineutrino accounting for the neutrino quantum decoherence ($\Gamma_1 = 10^{-21}$ GeV).

where $\beta$ represents the initial asymmetry between the electron and $l$-type neutrino ($l$ stands for the muon or tau neutrino), and $\bar{\beta}$ the asymmetry between electron and $l$-type antineutrinos, $\alpha$ is the ratio of electron antineutrinos relative to electron neutrinos, $n_e$ and $n_\nu$ describe the electron and neutrino density profiles, $G_F$ is the Fermi constant, $\theta$ is the neutrino mixing angle, $\delta m^2$ is the neutrino mass-squared difference.

We use for the numerical simulation the following initial conditions for the neutrino flux: $\alpha = 0.8$, $\beta = 0.48$ and $\bar{\beta} = 0.6$, and we consider neutrino energy is $E = 20$ MeV. The electron density profile is given by

$$n_e(r) = n_0 \left(\frac{R_\nu}{r}\right)^3 \left[a + b \tan^{-1}\left(\frac{r - R_\nu}{R_s}\right)\right], \quad (8)$$

where $R_\nu = 10$ km is the radius of the neutrinosphere, $n_0 = 10^{-4}$ eV is the electron density at the neutrinosphere, $a = 0.308$, $b = 0.121$ and $R_s = 42$ km are the parameters that characterise the electron fraction in the supernovae. The neutrino density profile is given by

$$n_\nu(r) = n_\nu^0 \left(\frac{R_\nu}{r}\right)^4, \qquad (9)$$

where $n_\nu^0 = 10^{-4}$ eV is the neutrino density at the neutrinosphere.

Within the numerical simulation of the neutrino evolution in supernovae environment we have obtained the survival probability $P_{\nu_e \nu_e}$ of the electron neutrino depending on the distance for two particular cases: (i) in the absence of the neutrino quantum decoherence (Fig. 1a), and (ii) for the case of the non-zero decoherence parameter $\Gamma_1 = 10^{-21}$ GeV (Fig. 1b). We have obtained the similar results for the survival probability $\overline{P}_{\nu_e \nu_e}$ of the electron antineutrino that is shown at Figs. 1c and 1d. One can see that the bipolar collective neutrino oscillations are suppressed by the quantum decoherence of the neutrino mass states. The obtained results of the numerical simulation are in a good agreement with the analytical conditions of the existence of the bipolar collective neutrino oscillations that were derived in [2].

This work is supported by the Russian Science Foundation under grant No.22-22-00384. K.S. acknowledges the support from the National Centre for Physics and Mathematics (Sarov, Russia) and the support from the Foundation for the Advancement of Theoretical Physics and Mathematics "BASIS" under Grant No. 20-2-2-3-1.